# Variability Type Determination and High Precision Ephemeris for NSVS 7606408

**Riccardo Furgoni**

*Keyhole Observatory, Via Fossamana 86, S. Giorgio di Mantova (MN), Italy; riccardo.furgoni@alice.it*

*and*

*Associazione Astrofili Mantovani—Gorgo Astronomical Observatory MPC 434, S. Benedetto Po (MN), Italy*

**Abstract** A photometric campaign analysis of the star NSVS 7606408 has been conducted in order to determine the type of variability and high accuracy ephemeris. By combining the obtained data with other datasets available, it was tried to improve the determination of the period, highlighting, however, a possible minimal change of the period over the years. At May 2, 2012, the ephemeris calculated for this variable star is HJDmin = $(2456015.44998 \pm 1.1 \times 10^{-4}) + E \times (0.35482573 \pm 3 \times 10^{-8}) + E^2 \times (2.4 \times 10^{-10} \pm 1 \times 10^{-11})$

## 1. Introduction

In recent years many wide field photometric surveys have led to the discovery of a large number of variable stars of different types. These include: NSVS (Northern Sky Variability Survey—an extensive variability survey of the sky north of Dec. −38° with daily time-sampling and a one-year baseline, by the Los Alamos National Laboratory; SuperWASP (Wide Angle Search for Planets)—UK's leading extra-solar planet detection program comprising a consortium of eight academic institutions, including Cambridge University, the Instituto de Astrofisica de Canarias, the Isaac Newton Group of telescopes, Keele University, Leicester University, the Open University, Queen's University Belfast, and St. Andrew's University; ASAS (All Sky Automated Survey)—a low-cost project dedicated to constant photometric monitoring of the whole available sky, which is approximately $10^7$ stars brighter than 14th magnitude—the project's ultimate goal is detection and investigation of any kind of photometric variability. Despite this, there has never been made a precise study of many of them with the aim of determining the class of variability and obtaining phase plots with a reasonably low scattering. Most of the instruments dedicated to large-scale research use short focal length telescopes which give extremely crowded images of star fields, making the photometric measurements difficult and sometimes inaccurate. The databases generated by these surveys, however, are a particularly valuable source of data in obtaining confirmations of newly discovered variable stars and in allowing a precise determination of the periods of regular variables. Having data concerning a large period of time makes period determination much more effective.

## 2. Instrumentation used

The data were obtained with a Celestron C8 Starbright, a Schmidt-Cassegrain optical configuration with aperture of 203 mm and central obstruction of 34%.

The telescope was positioned at coordinates: 45° 12' 32" N 10° 50' 20" E (WGS84), in a rural area with low to medium light pollution. The telescope was equipped with a focal reducer Baader Planetarium Alan Gee II able to bring the focal length from 2030 mm to 1396 mm. The focal ratio was reduced to $f/6.38$ from the original $f/10$.

The pointing was maintained with a Syntha NEQ6 mount with software SYNSCAN 3.27 guided through a Baader Vario Finder telescope equipped with a Barlow lens capable of bringing the focal length of the system to 636 mm and focal ratio of $f/10.5$.

The guide camera has been a Magzero MZ-5 with Micron MT9M001 monochrome sensor equipped with an array of $1280 \times 1024$ pixels. The size of the pixels is $5.2 \mu m \times 5.2 \mu m$ for a resulting sampling of 1.68 arcsec/pixel.

The CCD camera has been a SBIG ST8300m with monochrome sensor Kodak KAF8300 equipped with an array of $3352 \times 2532$ pixels. The pixels are provided with microlenses for improving the quantum efficiency, and the size of the pixels is $5.4 \mu m \times 5.4 \mu m$ for a resulting sampling of 0.80 arcsec/pixel. The camera has a resolution of 16 bits with a gain of 0.37e-/ADU and a full-well capacity of 25,500 electrons. The dark current is 0.02e-/pixel/sec at a temperature of –15° C. The typical read noise is 9.3e-RMS.

The camera is equipped with 1,000× antiblooming: after exhaustive testing it has been verified that the zone of linear response is between 1,000 and 20,000 ADU, although at up to 60,000 ADU the loss of linearity is less than 5%. The CCD is equipped with a single-stage Peltier cell $\Delta T = 35 \pm 0.1°$ C which allows cooling at a stationary temperature.

## 3. Data collection

The observations were performed with the CCD at a temperature of –5° C in binning $1 \times 1$. The exposure time was 55 seconds with a delay of 1 second between the images and an average download time of 11 seconds per frame. The star has been observed with a maximum pixel intensity of about 3,400 ADU and SNR 80. The comparison star had pixels with maximum intensity of about 4,000 ADU and SNR 100. The check star had pixels with maximum intensity of about 4,200 ADU and SNR 110. All stars have ADU intensity within the region of linear response of the CCD.

The observations were carried out without the use of photometric filters to maximize the signal-to-noise ratio. The spectral sensitivity of the CCD, as shown in Figure 1, is maximum at a wavelength of 540 nm, making the data more compatible with a magnitude CV (Clear Filter—Zero Point V).

The observations were conducted over eight nights as presented in Table 1.

The CCD control program was Software Bisque's CCD SOFT V5. Once the images were obtained, calibration frames were taken for a total of 30 dark of 55 sec at –5° C, 80 darkflat of 2 sec at –5° C, and 100 flat of 2 sec at –5° C. The darkflats and darks were taken only the first observation session and used for all other sessions. The flats were performed for each session as the position of the CCD camera could be varied slightly, as well as the focus point.

The calibration frames were combined with the method of the median and the masterframes obtained were then used for the correction of the images taken. All images were then aligned and an astrometric reduction was made to implement the astrometrical coordinate system WCS in the FITS header. These operations were conducted entirely through the use of software MAXIMDL V5.18 made by Diffraction Limited.

**4. The measurement of NSVS 7606408**

The star to be measured was observed within the Northern Sky Variability Survey (NSVS; Wozniak *et al.* 2004), which is a photometric survey for stars with an optical magnitude between 8 and 15.5. The research was conducted during the first generation of the Robotic Optical Transient Search Experiment (ROTSE-I) using a robotic system composed by four photographic lenses without photometric filters connected to CCDs. The research was conducted by Los Alamos National Laboratory (New Mexico) in order to cover the entire northern hemisphere of the sky. In a year of work between 1999 and 2000 an average of 150 photometric measurements for 14 million stars was made.

At the end of this research an automatically extended data analysis was made that led to the discovery of variable stellar sources. NSVS 7606408 has been recognized as a source with confirmed variability (Shaw *et al.* 2009). Identification data for this star are as follows:

| | |
|---|---|
| Name: | NSVS 7606408 |
| Position (UCAC3): | R.A. (2000), $11^h 48^m 56.586^s$, Dec. (2000) +26° 02' 30.07" |
| Cross Identifications: | GSC2.2 N201201146<br>CMC14 114856.5+260230<br>USNO-B1.0 1160-0187403<br>NOMAD1 1160-0192011<br>PPMX 114856.5+260230 |

UCAC3 233-104554
2MASS J11485658+2602301
SDSS J114856.58+260229.9

Proper motion (PPMX): pm R.A. = –0.08 mas/yr; pm Dec. = 2.39 mas/yr

AAVSO VSX period:   0.17740547 day (from Shaw *et al.* 2009)

The star field containing the variable star is shown in Figure 2.

The star's brightness was measured with MAXIMDL V5.18 software, using the aperture ring method. With a FWHM of the observing sessions at times arrived at values of 5" it was decided to choose values providing an adequate signal-to-noise ratio and the certainty of being able to properly contain the whole flux sent from the star. I have used the following apertures:

Aperture radius, 12 pixels; Gap width, 2 pixels; Annulus thickness, 8 pixels

The choice of making observations without the use of photometric filters required a proper study of the field in order to identify comparison stars with color similar to the variable under study, and with adequate signal-to-noise ratio, to obtain as accurate differential photometry as possible.

The choice was made by analyzing the measures relating to photometric 2MASS (Skrutskie *et al.* 2006) bands J, H, and K as mentioned in the *Carlsberg Meridian Catalog 14* (CMC14; Copenhagen Univ. Obs. *et al.* 2006). In the absence of precise measures of the magnitude in the standard V Johnson-Cousins, for stars used in the analysis it was decided to derive their V magnitude from the catalog CMC14 r' magnitude as described in Dymock and Miles (2009). This way ensures a good reliability and a final error theoretically not more than 0.05 magnitude in V. The conversion formula is as follows:

$$V_{\text{Johnson-Cousins}} = 0.6278 \times (J_{\text{2MASS}} - K_{\text{2MASS}}) + 0.9947 \times r'_{\text{CMC14}} \quad (9 < r' < 16) \quad (1)$$

The possible spectral type has been derived from the 2MASS J–H value instead, as shown in CMC14. The method is described in Stead and Hoare (2002) with the results as shown in Table 2.

Within the star field analyzed, only two stars have an adequate signal-to-noise ratio and color similar to the variable under study. Their position with respect to the variable is shown in Table 3 and Figure 3.

Through differential photometry the average magnitude of NSVS 7606408 is $13.572 \pm 0.006$ CV determined by placing the comparison star's magnitude at 13.325 CV. Whereas the variability of the star is by far superior to the gap that this value has with the V magnitude derived from CMC14 (= 13.641), it can be considered an acceptable result. The mean magnitude error for all the sessions is 0.006 CV. For the longest observing session the light curves of the variable star and the check star in relation to the comparison star are shown in Figure 4.

The check star shows an almost complete absence of significant trends: this fact shows that the color index of stars used is good enough for differential photometry carried out in unfiltered conditions.

## 5. Data analysis

Before proceeding further in the analysis, the time of the light curves obtained was heliocentrically corrected (HJD) in order to ensure a perfect compatibility of the data with observations carried out even at a considerable distance in time.

From here on, the obtained dataset will be referred to as FRIC in this work, (FRIC being the author's AAVSO observer code).

The determination of the period was calculated using the software PERIOD04 (Lenz and Breger 2005), using a Discrete Fourier Transform (DFT). The average zero-point (average magnitude of the object) was subtracted from the dataset to prevent the appearance of artifacts centered at a frequency 0.0 of the periodogram. All the data points were weighted in relation to the amount of the magnitude error. The periodogram was calculated with a frequency range from 0 to 25 $cd^{-1}$ (cycles per day) and step 0.00118537583 $cd^{-1}$. The result is shown in Figure 5.

The calculated frequency is 5.636659(89) $cd^{-1}$ (SNR 12.9) with an amplitude of $0.164 \pm 0.001$ CV. The period thus appears to be 0.1774100(28) day with a time of minimum 2456008.35376(17) HJD. The calculation of the uncertainties was carried out with PERIOD04 using the method described in Breger *et al.* (1999). The resulting phase diagram is shown in Figure 6.

In the VSX catalogue (Watson *et al.* 2012) maintained by the American Association of Variable Stars Observers (AAVSO), the star is listed as a possible eclipsing variable W UMa-type or as a pulsating variable δ Sct. By observing this phase diagram one can see that the curve shape much more resembles the rotation of an eclipsing variable rather than a pulsating one. In general, it is not easy to distinguish between EW eclipsing binaries and pulsating stars such as δ Sct or RRC since many of them have symmetric light curves, and sometimes comparable periods. In any case it is possible to distinguish them using the following criteria:

- The minima of the EW variables are generally sharper than maxima.

- The minima of the EW variables may have slight differences in depth.

- The EW variables can have maxima with slight differences in intensity before and after the primary minimum (O'Connell Effect) (Liu and Yang 2003).

- The RRC variables may have light curves slightly asymmetrical with the growth phase slightly steeper than the descending phase.

Based on these considerations and the more plausible classification of this variable as a W UMa system, I propose a new period for NSVS 7606408 equal to $P = 0.3548200(28)$ day and $HJD_{min} = 2456015.45016(17)$. The new phase diagram is shown in Figure 7. The low dispersion phase diagram allows us to recognize the following parameters:

- The minima centered at phase 0 and 0.5 are sharper than maxima present at phases 0.25 and 0.75.

- Primary minimum placed at phase 0 is deeper than the minimum placed at phase 0.5. The difference in intensity is estimated on the order of $0.028 \pm 0.005$ magnitude (the result is obtained by calculating the difference between the average value of the data places between phase 0.95 and phase 0.05 and those located between phase 0.45 and phase 0.55).

- The maximum placed at phase 0.25 is slightly less bright than the maximum placed at phase 0.75. The difference in intensity is estimated on the order of $0.013 \pm 0.005$ magnitude (the result is obtained by calculating the difference between the average value of the data places between phase 0.2 and phase 0.3 and those located between phase 0.7 and the phase 0.8).

The observations suggest that this system should be classified as a eclipsing variable W UMa (EW-type star) with a perceptible difference between primary and secondary minimum and the existence of a probable O'Connell effect.

## 6. Other photometric data for NSVS 7606408

As a series of photometric measurements are accurate and provided with low scatter, the temporal extension of them is a key requirement to allow an accurate determination of the period. The longer their extension, in fact, the more noticeable will be the effect that a minimum error in the determination of the period will lend to the phase diagram obtained. Even the phenomenon of aliasing tend to decrease if in the presence of temporal coverage sufficiently long and provided with adequate resolution. My observations give a phase plot with reduced dispersion; until now, no phase diagram has been published for NSVS 7606408, except that associated with NSVS data, which however, is plagued by a very noticeable scattering.

It was therefore decided to evaluate all the existing photometric observations for this star able to provide a reasonable amount of data to improve the period specified above.

The following surveys were taken into consideration:

• NSVS   The star is listed as NSVS 7606408. There are 233 observations made from April 5, 1999, to March 30, 2000, with a mean magnitude of $13.72 \pm 0.07$ (the original data are 275 points but are reduced to 233 after the rejection on the flag N Good Points). The data were obtained through the SkyDot service operated by Los Alamos National Laboratory (Sky Database for Objects in Time-Domain. Data at http://skydot.lanl.gov/nsvs/star.php?num=7606408&mask=32004). I have made a heliocentric correction of the data.

• SuperWASP   The star is listed as 1SWASP J114856.59+260230.1. There are 5,916 observations made from May 2, 2004, to May 17, 2008 with a mean magnitude of $13.73 \pm 0.04$ (Tamflux2). The data, provided by the Public SWASP Archive in the form of Data Fits, were extracted using fv software, only for the values Tmid, Tamflux2, and Tamflux2_Err. (Tmid = HJD Mean Time of Exposure from JD Reference which is 2453005.5; Tamflux2 = Original flux (Tamuz) corrected on the basis of a decorrelation technique highlighted by Collier Cameron et al. (2006). This flux is provided in microvegas and its conversion to magnitudes is given by the formula mag. = $15 - 2.5 \log(\text{Flux})$; Tamflux2_err = error calculated on flux Tamflux2.) The SuperWASP data are already heliocentrically corrected and even for the average time of exposure. The wide availability of data related to this dataset led me to make a selection on those that showed the Tamflux2_Err less than 0.05 magnitude. The number of observations used is thus down to 3,158 with a time span fom May 4, 2004, to May 16, 2007, and a mean magnitude of 13.706.

• ASAS   The star is listed as ASAS 114858+2602.5. Only data relating to ASAS- 3 (V-Magnitude) are present, not to ASAS-2 (I-Magnitude). There are 139 observations made from January 2, 2003, to March 2, 2009, with a mean magnitude of $13.67 \pm 0.03$. (The original data are 279 points but are reduced to 139 after discarding all those with flag D (worst data, probably useless) and all those having at least one measure invalid (Mag0; Mag1; Mag2; Mag3; Mag4 = 29.999. The average magnitude was calculated by performing for each given data the average of the values  Mag1, Mag2; Mag3; Mag4 corresponding to different apertures used for the calculation of the stellar flux.) The data were provided by the Astronomical Observatory, University of Warsaw, and are already heliocentrically corrected.

Table 4 shows the period determinations for each dataset using the software period04. (When the same criteria used for determining the period and uncertainties of FRIC dataset data were also used here, they have been omitted.) The respective phase diagrams and Fourier spectra are shown in Figures 8 through 13.

**7. Combining the available data**

Due to the different spectral sensitivity of the CCDs used in these surveys (as well as the lack or the use of different filters), the combination of the datasets can not be trusted as regards the determination of the amplitude variation. Using the DFT, only the SWASP dataset shows an amplitude of variation at all similar to the FRIC dataset (0.164±0.001) while all the other datasets show slightly different values. Moreover, the NSVS and ASAS datasets have a strong scattering which could make inaccurate the combined Fourier analysis. Since it is unclear how these differences affect the DFT of the combined data, only the FRIC and SWASP datasets will be used in combination for this calculation. In this case, the DFT may be used only to obtain a possible period to be used as a good starting point for a more detailed O–C analysis.

In order to combine the datasets it was first necessary to normalize the mean magnitude of the SWASP, NSVS, and ASAS datasets to the average value of the FRIC dataset. (For the NSVS and ASAS datasets the offset has been applied only for the realization of the combined phase diagram shown in Figure 15.) The applied offsets are given in Table 5.

The periodogram obtained, calculated on the frequency between 0 and 15 (cd$^{-1}$) with steps of 0.00001712275 (cd$^{-1}$) is shown in Figure 14.

The choice of combining the two datasets has led to a significant reduction in errors relating to the period, the amplitude variation and the determination of the time of primary minimum.

The calculated period is 5.63656978(8) cd$^{-1}$ (SNR=13.068), which corresponds to 2P=0.35482573(3) day, the amplitude is 0.1645±0.0007 mag. with HJD$_{min}$=2456015.44998(11). The resulting phase diagram for all the datasets based on these new parameters is given in Figure 15.

Visually we can see that the NSVS dataset, which is the earliest in time, is clearly not in phase with the FRIC and SWASP. Trying to slightly vary the period to improve the phase of NSVS we immediately lose the coincidence of other datasets. Now it would be necessary to use an O–C analysis even if the NSVS, ASAS, and SWASP datasets did not have a sampling rate that allows to verify precisely the times of minimum. Due to the fact that there are no observations in the literature regarding the time of minimum for this variable star we will proceed as follows: for each dataset will be taken as times of minimum the times in which the light intensity is less than magnitude 13.75. (The low numbers of ASAS data and their high dispersion renders them useless for this type of analysis. This dataset was not used for the creation of the O–C diagram.) Obviously, because of scattering, even this calculation will be noisy, but if a period change is present, the O–C diagram should show the typical shape (Figure 16).

Compared to the primary minimum observed on 2456015.44998(11) HJD, the earlier NSVS primary minimum that best represents the time shift with respect to the calculated ephemeris is out of phase by 0.042(1) day. Here it is important to remember that the vertical shape of the different minima in the graph is due solely to the uncertainty of the derived time of timimum related to the dataset's scattering. I have used for fitting a second order polynomial and the good result obtained suggests that we are faced with a progressive slowing of the period. Now, if we assume that the period is slowing down for a probable mass transfer between the two stars, is necessary to calculate how much each cycle must slow down to accumulate in 13,326 cycles a time shift of 0.042(1) day. (These are the complete cycles that the star has made since the earlier NSVS primary minimum until $HJD_{min} = 2456015.44998 \pm 0.00011$ HJD.)

It is necessary that the period increases $4.8 \times 10^{-10} \pm 1 \times 10^{-11}$ days at each cycle to be able to accumulate a time shift of 0.042(1) day in 13,326 cycles. As the calculation is based on very small quantities it is necessary to provide a double check: it is possible to evaluate the reliability of this prediction by looking at the data provided by PERIOD04 concerning the calculation of the period of each dataset.

If the period really increases by this small amount in each cycle it is necessary that in the 13,326 cycles which separate the two minima it is increased by $6.396 \times 10^{-6}$ days. The period calculated by PERIOD04 for the earlier NSVS dataset is 0.3548184(14) day and if we add to this the value $6.396 \times 10^{-6}$ days we obtain 0.3548248(14), which is very close to the period obtained by PERIOD04 for the FRIC and SWASP data combined as shown in Table 5.

In other words, the different length of the period found by the DFT in the measurements carried out in 1999 and 2000 by NSVS in comparison with those carried out more recently is very similar to the variation of the period calculated with the O–C diagram.

## 8. Conclusions

By combining the available data I have tried to classify the star NSVS 7606408, originally considered in the literature a possible W UMa- or δ Sct-type variable star with a period of 0.17740547 day. Analyzing the obtained data it is believed possible to identify the type of variability and a precise calculation of the ephemeris. The final resulting data are as follows:

| | |
|---|---|
| Star name | NSVS 7606408 |
| Equatorial coordinates (UCAC3) | R. A. (2000) $11^h \, 48^m \, 56.586^s$ |
| | Dec. (2000) +26° 02' 30.07" |
| Variability type | Eclipsing binary W UMa |
| Actual Period | $0.354825732 \pm 3 \times 10^{-8}$ day |
| Amplitude variation (primary min – primary max) | |

$$13.743 \pm 0.005\,CV - 13.403 \pm 0.005\,CV = (0.340 \pm 0.005\,CV)$$

Difference between primary and secondary minimum
$$13.743 \pm 0.005\,CV - 13.715 \pm 0.005\,CV = (0.028 \pm 0.005\,CV)$$

Difference between primary and secondary maximum
$$13.403 \pm 0.005\,CV - 13.416 \pm 0.005\,CV = (0.013 \pm 0.005\,CV)$$

| | |
|---|---|
| Effects present | O'Connell Effect |
| Ephemeris | $HJD_{min} = (2456015.44998 \pm 1.1 \times 10^{-4})$ $+ E \times (0.35482573 \pm 3 \times 10^{-8})$ $+ E^2 \times (2.4 \times 10^{-10} \pm 1 \times 10^{-11})$ |

## 9. Aknowledgements

I wish to thank Mrs. Antonella Finotti for her support in revising the text, and Stefano Brangani, Ph.D., for his valuable comments on the mathematical aspects of this work.

Table 1. Activity log for the data collection concerning the FRIC dataset. The NExT value indicates the Net Exposure Time which is the session duration spent in pure photons collection.

| Date (dd-mm-yyyy) | UTC Start (hh:mm:ss) | UTC End (hh-mm-ss) | NExT[1] (hh-mm-ss) | Useful Exposures | Airmass Min | Airmass Max |
|---|---|---|---|---|---|---|
| 21-03-2012 | 18:41:11 | 22:34:41 | 02:55:07 | 202 | 1.065 | 1.767 |
| 22-03-2012 | 18:35:15 | 21:07:47 | 01:54:24 | 135 | 1.153 | 1.783 |
| 25-03-2012 | 20:57:53 | 22:44:12 | 01:19:44 | 94 | 1.058 | 1.149 |
| 26-03-2012 | 19:11:22 | 00:49:43[2] | 04:13:38 | 294 | 1.058 | 1.459 |
| 28-03-2012 | 20:08:08 | 00:45:15[3] | 03:27:50 | 240 | 1.058 | 1.230 |
| 22-04-2012 | 19:24:59 | 21:29:28 | 01:39:55 | 109 | 1.058 | 1.122 |
| 27-04-2012 | 20:38:54 | 23:29:46 | 02:17:30 | 150 | 1.058 | 1.282 |
| 02-05-2012 | 19:38:20 | 23:04:29 | 02:42:15 | 177 | 1.058 | 1.266 |

[1] *Net Exposure Time: session duration spent in photon collection.* [2] *00:49:43 of 27-03-2012.*
[3] *00:45:15 of 29-03-2012.*

Table 2. 2MASS Colors and Spectral Type relation as described in Stead and Hoare (2002).

| Spectral Type | No. stars per subtype | H–K (UKIDSS) | J–H (UKIDSS) | H–K (2MASS) | J–H (2MASS) | e(H–K) | e(J–H) |
|---|---|---|---|---|---|---|---|
| Main sequence intrinsic colors O8V to M0V | | | | | | | |
| O8V | 6 | –0.10 | –0.19 | –0.10 | –0.21 | 0.017 | 0.019 |
| O9V | 23 | –0.10 | –0.15 | –0.09 | –0.17 | 0.009 | 0.011 |
| B0V | 24 | –0.10 | –0.17 | –0.09 | –0.19 | 0.008 | 0.010 |
| B1V | 54 | –0.09 | –0.16 | –0.08 | –0.17 | 0.006 | 0.008 |
| B2V | 105 | –0.06 | –0.14 | –0.06 | –0.15 | 0.004 | 0.005 |
| B3V | 101 | –0.05 | –0.12 | –0.05 | –0.13 | 0.004 | 0.005 |
| B4V | 35 | –0.04 | –0.10 | –0.04 | –0.11 | 0.007 | 0.008 |
| B5V | 125 | –0.03 | –0.10 | –0.03 | –0.11 | 0.004 | 0.004 |
| B6V | 70 | –0.02 | –0.09 | –0.02 | –0.10 | 0.006 | 0.006 |
| B7V | 85 | –0.02 | –0.08 | –0.02 | –0.09 | 0.005 | 0.006 |
| B8V | 240 | –0.02 | –0.08 | –0.01 | –0.09 | 0.003 | 0.003 |
| B9V | 351 | 0.00 | –0.06 | 0.00 | –0.06 | 0.002 | 0.003 |
| A0V | 195 | 0.00 | –0.04 | 0.00 | –0.04 | 0.003 | 0.004 |
| A2V | 353 | 0.02 | –0.02 | 0.02 | –0.02 | 0.002 | 0.003 |
| A5V | 215 | 0.03 | 0.02 | 0.03 | 0.03 | 0.003 | 0.004 |
| F0V | 232 | 0.04 | 0.09 | 0.04 | 0.10 | 0.003 | 0.004 |
| F2V | 245 | 0.03 | 0.13 | 0.04 | 0.14 | 0.003 | 0.004 |
| F5V | 231 | 0.05 | 0.16 | 0.05 | 0.18 | 0.003 | 0.004 |

Table 2. 2MASS Colors and Spectral Type relation as described in Stead and Hoare (2002), cont.

| Spectral Type | No. stars per subtype | H–K (UKIDSS) | J–H (UKIDSS) | H–K (2MASS) | J–H (2MASS) | e(H–K) | e(J–H) |
|---|---|---|---|---|---|---|---|
| F8V | 146 | 0.05 | 0.19 | 0.06 | 0.21 | 0.004 | 0.005 |
| G0V | 115 | 0.06 | 0.22 | 0.07 | 0.25 | 0.005 | 0.006 |
| G2V | 95 | 0.06 | 0.24 | 0.07 | 0.27 | 0.005 | 0.006 |
| G5V | 141 | 0.06 | 0.28 | 0.07 | 0.31 | 0.005 | 0.006 |
| G8V | 171 | 0.06 | 0.35 | 0.07 | 0.38 | 0.004 | 0.005 |
| K0V | 136 | 0.07 | 0.37 | 0.08 | 0.40 | 0.005 | 0.006 |
| K2V | 47 | 0.08 | 0.40 | 0.09 | 0.44 | 0.008 | 0.010 |
| K5V | 20 | 0.12 | 0.50 | 0.13 | 0.55 | 0.014 | 0.019 |
| M0V | 12 | 0.19 | 0.60 | 0.21 | 0.67 | 0.015 | 0.019 |

Giant intrinsic colors: G4III to M5III.

| Spectral Type | No. stars per subtype | H–K (UKIDSS) | J–H (UKIDSS) | H–K (2MASS) | J–H (2MASS) | e(H–K) | e(J–H) |
|---|---|---|---|---|---|---|---|
| G5III | 156 | 0.127 | 0.436 | 0.188 | 0.465 | 0.008 | 0.008 |
| G8III | 585 | 0.128 | 0.458 | 0.119 | 0.487 | 0.004 | 0.005 |
| K0III | 514 | 0.132 | 0.493 | 0.123 | 0.522 | 0.004 | 0.005 |
| K2III | 578 | 0.174 | 0.581 | 0.166 | 0.613 | 0.006 | 0.006 |
| K5III | 426 | 0.223 | 0.778 | 0.215 | 0.814 | 0.014 | 0.014 |
| M0III | 179 | 0.237 | 0.808 | 0.230 | 0.845 | 0.023 | 0.022 |
| M2III | 353 | 0.243 | 0.903 | 0.235 | 0.941 | 0.017 | 0.017 |
| M5III | 73 | 0.327 | 0.912 | 0.320 | 0.956 | 0.041 | 0.041 |

Table 3. Comparison star positions and information with respect to NSVS 7606408.

| Object | R.A. (2000) UCAC3 h m s | Dec. (2000) UCAC3 ° ' " | 2MASS J–H | CMC14 r'mag | CMC14 derived V mag. | Probable spectral type |
|---|---|---|---|---|---|---|
| NSVS7606408 | 11 48 56.586 | +26 02 30.07 | 0.226 | 13.532 | 13.641 | F8V-G0V |
| Comp. Star[1] | 11 49 09.986 | +25 55 32.92 | 0.229 | 13.192 | 13.325 | F8V-G0V |
| Check Star[2] | 11 49 37.454 | +26 14 48.23 | 0.208 | 13.145 | 13.275 | F5V-F8V |

[1] Cross references for comparison star: GSC2.2 N201201182; CMC14 114909.9+255532; USNO-B1.0 1159-0186938; NOMAD1 1159-0191730; PPMX 114909.9+255532; UCAC3 232-105102; 2MASS J11490997+2555328; SDSS J114909.99+255532.8

[2] Cross references for check star: GSC2.2 N20120112; CMC14 114937.4+261448; USNO-B1.0 1162-0194828; NOMAD1 1162-0199537; PPMX 114937.4+261448; UCAC3 233-104578; 2MASS J11493745+2614483; SDSS J114937.45+261448.2

Table 4. Period determination results for each dataset, using PERIOD04 software.

| Dataset | Checked Range | Step | SNR | Period (days) | Amplitude |
|---------|---------------|------|-----|---------------|-----------|
| NSVS | 0–25 (cd$^{-1}$) | 0.00013895 (cd$^{-1}$) | 7.4 | 0.3548184 (14) | 0.170 ± 0.005 |
| SWASP | 0–25 (cd$^{-1}$) | 0.0000452 (cd$^{-1}$) | 11.9 | 0.35482227 (13) | 0.164 ± 0.001 |
| ASAS | 5–6 (cd$^{-1}$) | 0.0000222 (cd$^{-1}$)* | 4.0 | 0.3548227 (14) | 0.157 ± 0.027 |

*Several problems afflicting this dataset: high scattering, a few observations available. Despite this, forcing the search for a frequency only between 5 (cd$^{-1}$) and 6 (cd$^{-1}$) we obtain a result consistent with other datasets.

Table 5. Offsets applied to the mean magnitude of each dataset used.

| Dataset | Original mean mag. | Applied offset | New mean mag. |
|---------|--------------------|-----------------|---------------|
| FRIC (Furgoni, R.) | 13.572 | — | — |
| SWASP | 13.73 | –0.158 | 13.572 |
| NSVS | 13.72 | –0.148 | 13.572 |
| ASAS | 13.67 | –0.098 | 13.572 |

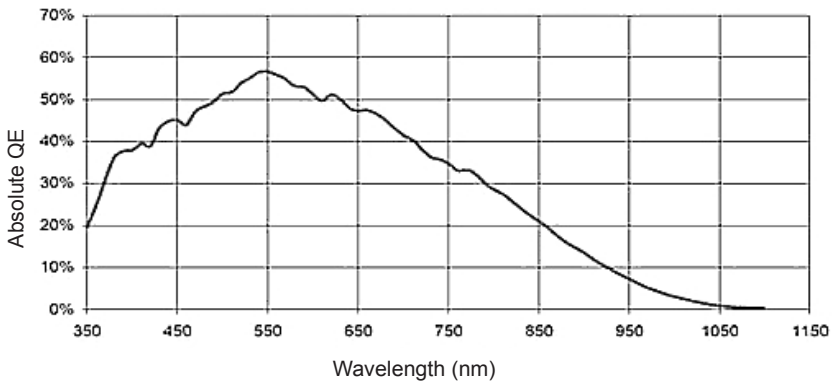

Figure 1. Quantum Efficiency of the Kodak KAF8300 monochrome sensor with microlens, equipping the CCD SBIG ST8300m used for the data collection of this study.

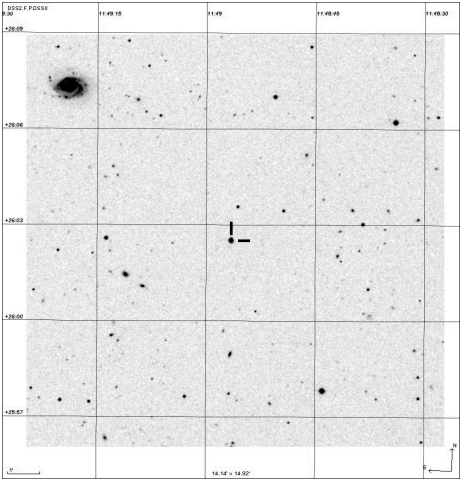

Figure 2. DSS image centered on NSVS 7606408. In the top-left side of the field the galaxy NGC 3902 is clearly visible. North is up, East to the left.

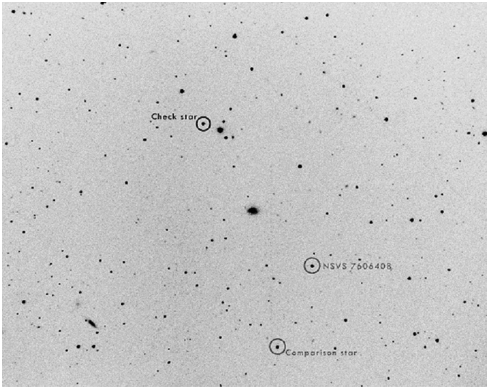

Figure 3. Star field containing NSVS 7606408, the comparison star UCAC3 232-105102 and the check star UCAC3 233-104578. North is up, East to the left.

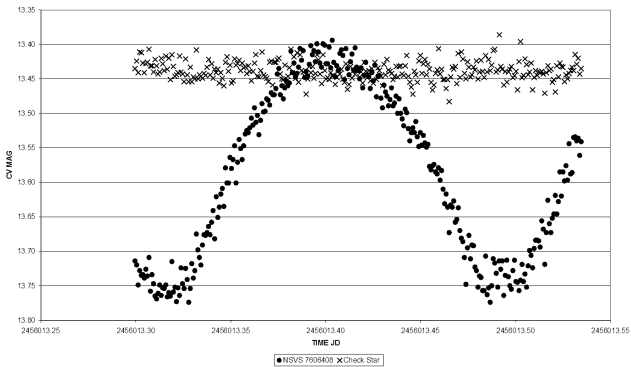

Figure 4. Differential photometry of NSVS 7606408 and the check star UCAC3 233-104578 in relation to the comparison star UCAC3 232-105102. The 26/03/2012 session is shown, which is the longest one. The check star shows an almost complete absence of significant trends: this fact shows that the color index of stars used is good enough for differential photometry carried out in unfiltered conditions.

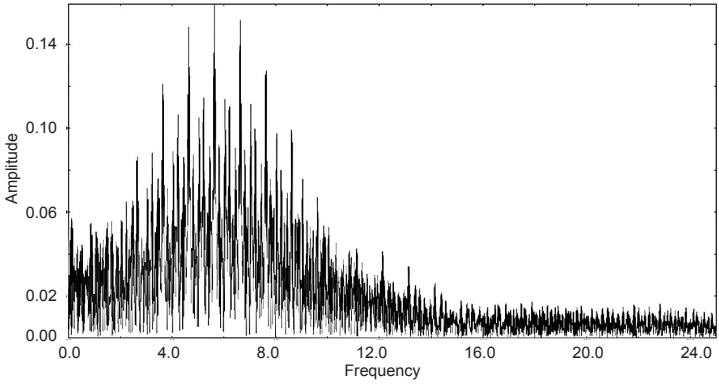

Figure 5. Fourier spectrum of the FRIC dataset weighted in relation to the magnitude error. The periodogram was calculated with a frequency range from 0 to 25 $cd^{-1}$ (cycles per day) and step 0.00118537583 $cd^{-1}$.

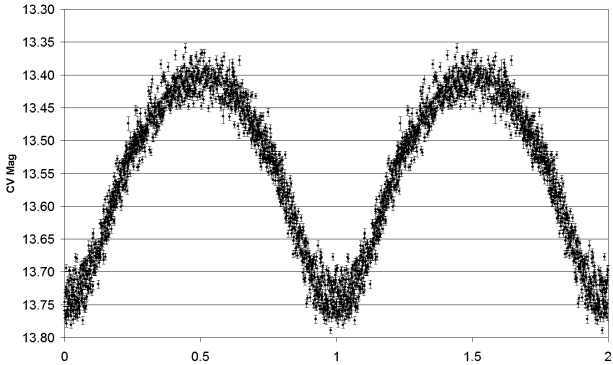

Figure 6. Phase plot of the FRIC dataset with period 0.1774100 day.

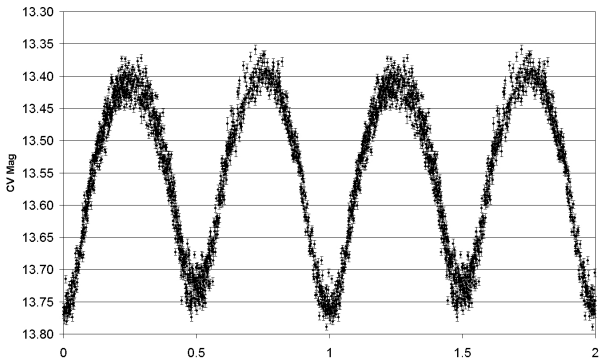

Figure 7. Phase plot of the FRIC dataset with period 0.3548200 day. The shape of a W Uma type variable star is evident.

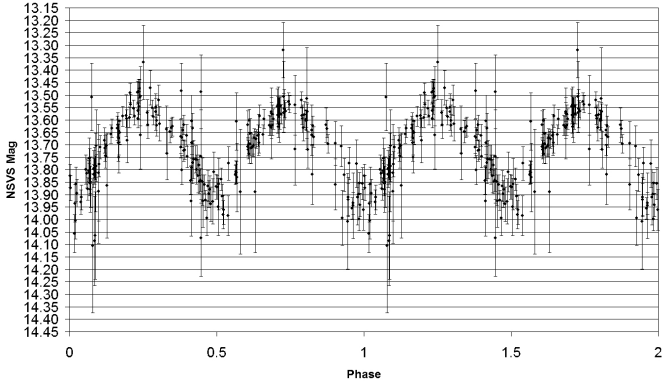

Figure 8. Phase plot of the NSVS dataset with period 0.3548184 day.

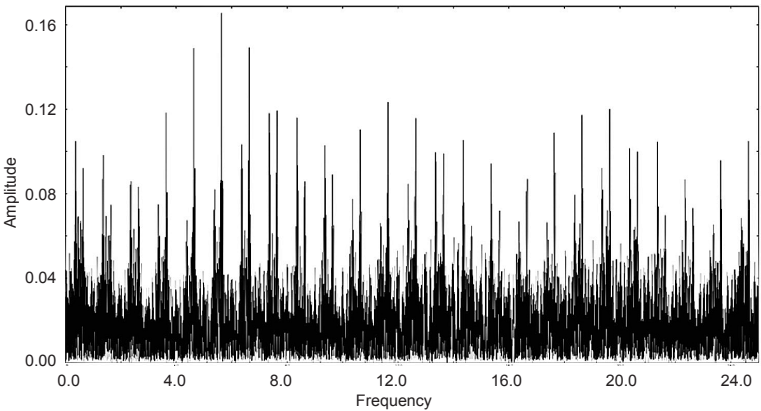

Figure 9. Fourier spectrum of the NSVS dataset weighted in relation to the magnitude error. The periodogram was calculated with a frequency range from 0 to 25 cd$^{-1}$ (cycles per day) and step 0.00013895 (cd$^{-1}$).

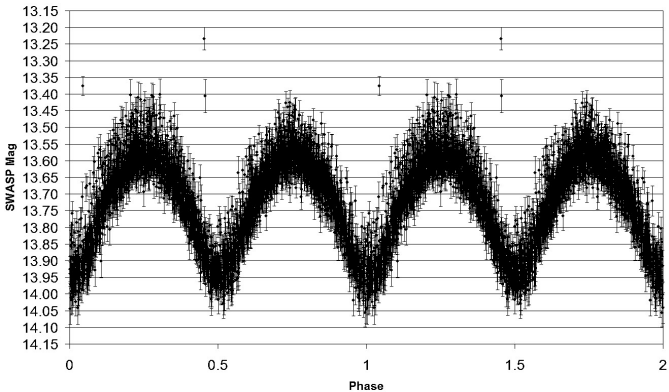

Figure 10. Phase plot of the SWASP dataset with period 0.35482227 day.

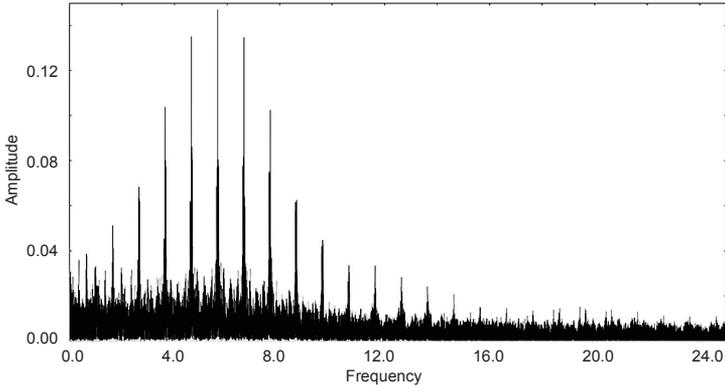

Figure 11. Fourier spectrum of the SWASP dataset weighted in relation to the magnitude error. The periodogram was calculated with a frequency range from 0 to 25 cd$^{-1}$ (cycles per day) and step 0.0000452 (cd$^{-1}$).

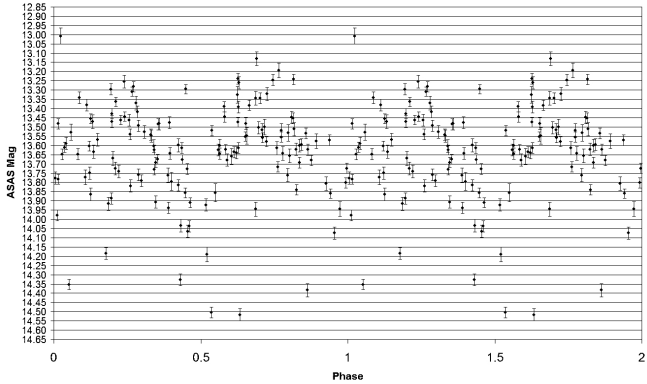

Figure 12. Phase plot of the ASAS dataset with period 0.3548227 day.

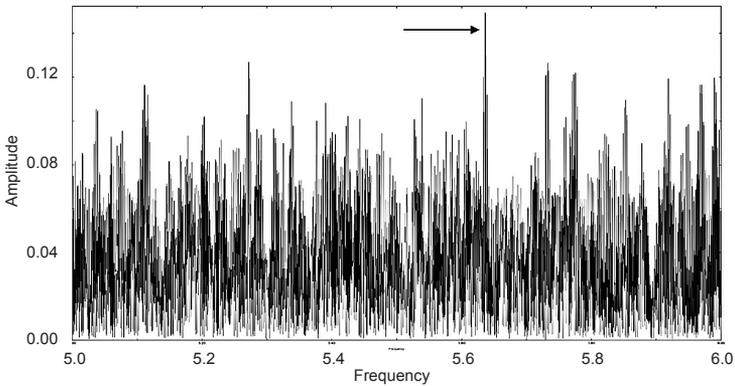

Figure 13. Fourier spectrum of the ASAS dataset weighted in relation to the magnitude error. The periodogram was calculated with a frequency range from 0 to 25 cd$^{-1}$ (cycles per day) and step 0.0000222 (cd$^{-1}$). The arrow marks the significant peak.

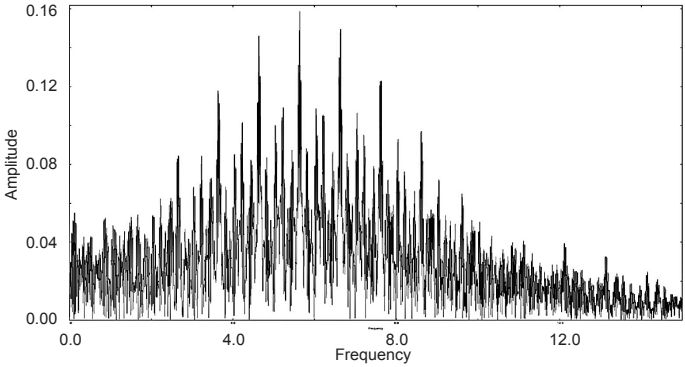

Figure 14. Fourier spectrum of the FRIC + SWASP datasets. An offset of –0.158 mag has been applied to SWASP to normalize the mean magnitude to that of FRIC. The periodogram was calculated with a frequency range from 5 to 6 cd$^{-1}$ (cycles per day) and step 0.00001712275 (cd$^{-1}$).

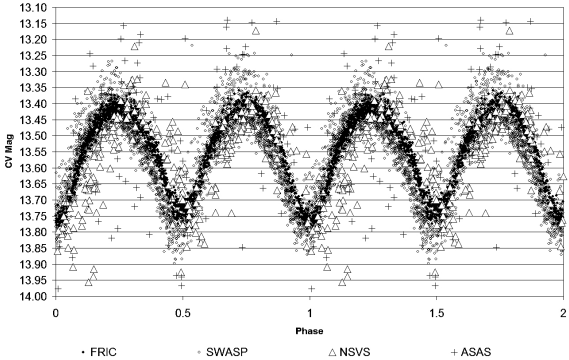

Figure 15. Phase plot of FRIC, SWASP, NSVS, and ASAS dataset with period of 0.35482573 day. We can see that the NSVS dataset, which is the earliest in time, is clearly not in phase with the FRIC and SWASP.

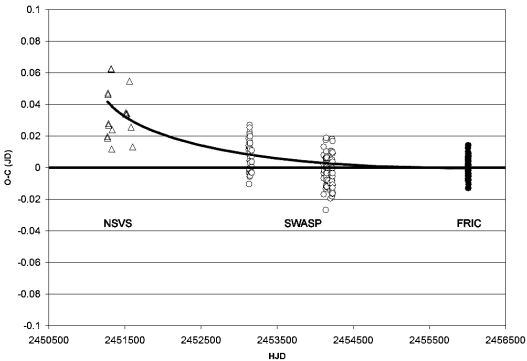

Figure 16. O–C analysis for the time of primary minimum of NSVS 7606408.